\documentclass[a4paper,UKenglish,cleveref, autoref, thm-restate]{lipics-v2021}
\pdfoutput=1 

\hideLIPIcs  

\title{Generalize Synchronization Mechanism: Specification, Properties, Limits} 

\author{Chih-Wei Chien}{National Cheng Kung University, Taiwan R.O.C. \and \url{https://github.com/idoleat} }{chiencw@taiker.tw}{}{}

\author{Chi-Yeh Chen}{National Cheng Kung University, Taiwan R.O.C.}{chency@csie.ncku.edu.tw}{}{}

\authorrunning{C.\,W. Chien and C.\,Y. Chen} 

\Copyright{Chih-Wei Chien and Chi-Yeh Chen} 



\category{} 

\relatedversion{} 

\EventEditors{John Q. Open and Joan R. Access}
\EventNoEds{2}
\EventLongTitle{42nd Conference on Very Important Topics (CVIT 2016)}
\EventShortTitle{CVIT 2016}
\EventAcronym{CVIT}
\EventYear{2016}
\EventDate{December 24--27, 2016}
\EventLocation{Little Whinging, United Kingdom}
\EventLogo{}
\SeriesVolume{42}
\ArticleNo{23}

\usepackage{tikz}
\usetikzlibrary{arrows,shapes,automata,petri,positioning,calc}
\usetikzlibrary{arrows,automata,positioning}
\usetikzlibrary{arrows,shapes}
\tikzset{
    place/.style={
        circle,
        thick,
        draw=black,
        fill=gray!50,
        minimum size=6mm,
    },
        state/.style={
        circle,
        thick,
        draw=blue!75,
        fill=blue!20,
        minimum size=6mm,
    },
}
\usepackage{dsfont}
\newcommand\two[2]{\renewcommand{\arraystretch}{1.2}\setlength\tabcolsep{0pt}\begin{tabular}[c]{l}#1\\#2\end{tabular}}
\nolinenumbers

\begin{document}

\maketitle

\begin{abstract}
Shared resources synchronization is a well studied problem, in both shared memory environment or distributed memory environment. Many synchronization mechanisms are proposed, with their own way to reach certain consistency level. This thesis further found that there is no perfect synchronization mechanism. Each of them has its properties at different level. For example, to enforce strong consistency, writers may loose writing freedom or it would take more time to coordinate. This thesis proposes a framework to generalize all synchronization mechanism in a formal way for better reasoning on properties, from the perspective of multi-writer to single-writer convergence. Therefore, limitations prevent a synchronization mechanism from achieving every property at its optimal level. CAP and ROLL were proposed in previous works to explain such. CAP theorem states that it can only achieve two of Consistency, Availability and Partition tolerance properties. ROLL Theorem uses a framework to model leaderless SMR protocol and states quorum size and fault tolerance are trading off. The thesis covers five properties in a more understandable way to analyze trade-offs and explore new mechanisms.
\end{abstract}

\section{Introduction} \label{ch:1-introduction}

To achieve speedup, scalability or even fault tolerance, we often leverage multiple workers such as threads or processes operating on the same task. In ideal scenario that the program is Data-Race-Free, the goal can be achieved easily since we have data parallelism. Data-intensive applications such as image/audio processing or scientific computing often fall in this category. In scenarios involving the utilization of shared resources, such as sharing a variable, a data structure or a sequence of log, it is imperative to use synchronization mechanisms to coordinate workers for correct and expected results.

Many synchronization mechanisms have been proposed to solve this problem. In shared memory environment, notably locks, semaphores, RCU, etc. are common ones. In distributed memory environment, we can keep using the mechanisms in shared memory environment by adapting remote direct memory access, building an illusion of shared memory. The other way is identifying each piece of distributed memory as a replication, making them shared nothing. Famous algorithms on synchronizing each replication are notably Paxos family\cite{10.1145/279227.279229}, Raft\cite{184040}, Viewstamp Replication\cite{10.1145/62546.62549}, Leaderless SMR family\cite{rezende2020leaderless} and CRDTs\cite{10.1007/978-3-642-24550-3_29}.

Although every synchronization mechanism can eventually make workers synchronized on shared resources, each of them has different properties. From strong consistency, low writing freedom to weaker consistency, high writing freedom, alone side with other properties such as loading and fault tolerance. We can observe that there is no perfect synchronization mechanism that can hold all the properties. Properties can trade off each other. To the best of our acknowledgment, there is only a previous work ROLL theorem\cite{rezende2020leaderless} proposed a framework to model and specify leaderless SMR protocols, reasoning about how loading and fault-tolerance trading off. Now we try to expand the research into modeling all the synchronization mechanisms in a formal way, in order to analyze the properties and limitations.

In this thesis, we propose a framework to model synchronization mechanisms. The framework first generalizes synchronization mechanisms as multi-writer to single-writer convergence. Then, analyze how mechanisms converging in different ways at different stage, with different quorum. It brings out the properties of each mechanism. By using this framework we can analyze and reason about how mechanisms converge, capturing how good or how bad mechanisms are on each property. We can further found limits of mechanisms on trading off properties. If an mechanism is right at the limit we say it is optimal. By placing all the possible analysis, we found there is still some room for new way of synchronization.

\section{Related Works} \label{ch:2-related works}

Although many algorithms on synchronization has been proposed, the study on generalizing synchronization mechanism and analyzing properties has not been well studied.

In Leaderless State-Machine Replication: Specification, Properties, Limits\cite{rezende2020leaderless}, Tuanir França Rezende and Pierre Sutra proposed a framework to decompose leaderless state-machine replication protocols. By using the framework to represent leaderless SMR protocols, the ROLL properties (Reliability, Optimal Latency and Load balancing) on leaderless SMR protocols are defined. Further more, protocols with these properties satisfy the ROLL Theorem: $2F + f - 1 \leq n$, where $n$ is the total process number, $n-F$ is the quorum size, f is the maximum number of failures can be tolerated. \cite{rezende2020leaderless} provides the high level view of What leaderless SMR protocols are and how they behave in a formal way, with an inequality stating the trade-off between quorum size and fault tolerance.

By using the framework, we can have a clear way to reason about how leaderless SMR protocols or even more synchronization mechanisms work. The framework can help analyze properties on synchronization mechanisms and check if it reaches the limit or what property to sacrifice for gaining more on another. When developers or researchers want to design a new synchronization mechanism or improve current ones, the framework could be an useful template or reference.

With the similar goal, we propose a framework but aiming at cover synchronization mechanisms in general, from the perspective of multi-writer to single-writer convergence. By modeling and analyzing synchronization mechanisms with the framework, it captures five key properties: consistency, writing freedom, latency, loading and fault tolerance. Further limits on trading off properties are presented by the framework as well, including the ROLL theorem from \cite{rezende2020leaderless}. Comparing to \cite{rezende2020leaderless}, we start from a different perspective and our framework is more universal on modeling synchronization mechanisms by including mechanisms on both distributed memory and shared memory with different consistency. Also our framework is simpler and more understandable than the one in \cite{rezende2020leaderless}. In the work In Search of an Understandable Consensus Algorithm\cite{184040}, it shows that simplicity matters.



\section{Synchronization-Framework}

\subsection{System Model}
In this section we describe the environment in the following discussion. In a shared memory environment, a register $r$ is an infinite series of memory address $S = \{a_1, a_2,...\}$. A writer performs converged $write$ operation to a register by writing value in a new address and appending the address to the end of the series. For example, $S_i$ is a series with size $i$. With the $write$ operation writing a converged value at $a_{i+1}$, $S_{i+1} = \{S_i, a_{i+1}\}$. Readers perform $read$ by dereferencing value at each address ($^*a$) in the series through to the latest possible address and applying projecting action $\mathscr{F}$. Projection result is then returned by $\mathscr{F}(S_i)$. $\mathscr{F}$ could be simply taking the latest one: $\mathscr{F}(S_i) = {^*a_i}$, summarizing all the value up: $\mathscr{F}(S_i) = \sum_{n=1}^{i} {^*a_n}$ or using other rules. Since readers will not conflict with writers (will not operate on the same address), the rest of this thesis will only focus on writers. In a distributed memory environment, if a shared memory illusion is created through remote memory access, a register is defined the same as above. If each distributed memory is a piece of replication, each replication is considered as an individual register, defined in the same way as well. The register can implement common resources to be synchronized, such as a variable, a counter or a replicated log, by using different projecting action $\mathscr{F}$.

A valid synchronization mechanism can synchronize all the registers to the same projection result. To achieve this, a valid synchronization mechanism must show correctness and progress. 
\begin{itemize}
    \item \textbf{Correctness:} The mechanism success with the same projection result on all registers.
    \item \textbf{Progress:} A synchronization mechanism will eventually terminate.
\end{itemize}
An invalid synchronization mechanism may end up with wrong result or stuck at some point (without any progression). The projection results of all registers would not be the same.

The framework we are going to propose works on both multiple writers on one register and multiple writers across multiple registers (one on each). On failure mode, a writer may crash and recover. Byzantine fault is not considered.

\subsection{Synchronization Is a Process of Multi-writer to Single-writer Convergence} \label{ch3.2}

Before introducing the framework, we first explain how synchronization mechanisms work in general. In distributed memory environment, concurrent $write$s on different registers could lead to different states. Concurrent $write$s are the $write$ operations that do not have happens-before relationship. Even if every replication has the acknowledge of each $write$s, conflicts occur due to not knowing the order to apply $write$s. Each register still could end up with different states caused by different execution order, where number of states may grow pretty quick into state explosion. Synchronization mechanisms solve this by giving $write$s order or making sure $write$s can be commutative by design to get the same projection results on replications. Number of states are then shrunk down. Depending on the design of the mechanisms, orders could be partial order or total order. This is a process of Multi-writer to Single-writer convergence. Each successful pass of synchronization can be seen as a write operation which contains one or more values across all registers. More details are described in the framework (\textsection \ref{ch3.4}).

In shared memory environment, including distributed memory environment with shared memory illusion, multiple writers act on the same register. Concurrent $write$s without mutual exclusion cause race condition. When two writers try to append a memory location to the sequence, the latest location may have undesired result if the operation is not performed atomically. In C Language 2011 Standard, such behaviors are undefined \cite{ISO:2011:IIP}. Current synchronization primitives solve this by only allowing one writer entering the critical section. This is also a process of Multi-writer to Single-writer convergence. Even with data race avoided, concurrent $write$s still have the execution order problem described in the above distributed memory environment. Thus, synchronization mechanisms are applied to reach Multi-writer to Single-writer convergence.

In both shared memory and distributed memory environment, we can observe that synchronization mechanisms are designed around two core aspects: (1) The spread of fact. (2) What to do after receiving the fact. These two aspects shapes the form of how each synchronization mechanism converges. The following section will reason about some existed synchronization mechanisms from the perspective of multi-writer to single-writer convergence. 

\subsubsection{Examples of Convergence} \label{ch3.2.1}

From the perspective of the Multi-writer to Single-writer convergence, we attempt to generalize synchronization mechanisms and analyze core properties on them. In this section we will focus on explaining convergence by examining facts of $write$. Properties will be presented in \textsection \ref{ch3.3}. The thesis will then introduce a framework in \textsection \ref{ch3.4} to model and analyze synchronization mechanisms in \textsection \ref{ch3.5}.

\subparagraph*{Paxos.} In a distributed memory environment, each peer in Paxos can issue $write$ operation. Paxos converges to only one writer by two steps. At least the majority of peers grant one peer the permission to write, then accepting $write$ with the newest permission. Since each peer only accept the $write$ higher than the last accepted one, outdated writers (even with the permission) would become readers to learn the fact that it is unable to write. The accept message contains the event ID of the last accepted $write$, so every converged writer knows it bases on which previous $write$. All facts are linearized as a linked list conceptually.

Noting that the Paxos here includes the chaining use of Paxos instances, which is known as a kind of multi-Paxos implementation, but not the leader-based multi-Paxos.

\subparagraph*{Raft.} In a distributed memory environment, Raft elects a leader periodically to be the only one who can perform $write$. All writers converge to the only one leader before any operation. Facts easily get ordered since only one can write. Each $write$s are guaranteed to be replicated to each replication, so all the facts are still linearized if a new leader is elected. Writers in the leader-based multi-Paxos converge in the same way as well.

\subparagraph*{Viewstamp Replication.} Same as Raft, Viewstamp Replication takes a leader-based approach. Instead of electing a leader, Viewstamp Replication assigns each peer an index and iterates the index in round-robin way to rotate the leader role. A monotonically increased viewstamp number is used to record the current view on index. Writers are converged inherently since each peer knows how to calculate the index and current view. Communications are only needed when failures happen. All facts are dealt by the leader and view changes are guaranteed to converge to the same index.

\subparagraph*{EPaxos.} It is a Paxos variant. Similar to Paxos, each peer can issue $write$ operation, but without the need to gain permission. If there is no conflict, the value will be accepted directly, otherwise the $write$ will be rejected. The returned conflicting messages contain dependencies need to be committed before performing $write$ again. The fact of $write$s and whether they are conflicting or not is determined by the peers that receive those $write$s. All $write$s converge at those peers to the one that is based on the latest $write$s. Facts are linearized as long as the fact of each $write$ (dependencies) are passed along on every operation. EPaxos also prevents minor failures by having the majority having the fact by having larger quorum.


\subparagraph*{CRDTs.} All the above synchronization mechanisms aim for strongest consistency, linearizability. History is linearized at least in one member's eyes, for non-commutative operations. Conflict-free Replicated Data Type is designed to have commutative operations on the shared resources by leveraging monotonicity. Data may be inconsistent across the replicas, but eventually all the replicas will have the same projection result if they have received all the operations. CRDTs has weaker consistency but free from taking round trips on forcing ordering. All the $write$s converge to a set of $write$s, which can be seen as a writer containing multiple value. In other words, a set of $write$s is a batch $write$. CRDTs guarantee the projection result will consistent in a batch $write$, or strong eventual consistency in a more formal term. Once the fact of each $write$ is delivered 

\subparagraph*{Atomic operations.} In shared memory environment, atomic operations guarantee the correctness and more importantly the progression, by ensuring the converged $write$ operation is based on the previous converged $write$ operation. On each operation, a writer will read, modify and write as a whole. The operation is successful if no any other operation finished in between the whole read, modify, write. Other wise the operation is considered failed. Among concurrent $write$s, the converged one is the successful writer. All the other failed writers then get the fact that it has converged to another writer, which also indicates progression guarantee.





\subsection{Properties in Synchronization Mechanisms} \label{ch3.3}

Different applications and scenarios have varying constraints, which is why synchronization mechanisms offer different properties to address specific requirements and trade-offs. For instance, a real-time database system may prioritize low latency and strong consistency, while a distributed file system might focus on fault tolerance and scalability. It is crucial to choose the appropriate synchronization mechanism based on the specific needs of the system to achieve the desired balance among properties. 

A valid pass of synchronization is having writers converged and having enough registers learned the value to ensure safety. By observing, we found these five core properties are the ones that most systems care about in a pass of synchronization: consistency, latency, writing freedom, loading and fault tolerance. 

\subsubsection{Consistency}
In consistency notation given by Paolo Viotti and Marko Vukolic \cite{10.1145/2926965}, different strength of consistencies are described as the order guarantee of the result from $read$. For example in linearizability, $read$s are guaranteed to retrieve the newer value that is written by a newer $write$ by any writer globally. It is commonly used in distributed storage system to ensure correctness. In the weaker one, sequential consistency, $read$ order is only guaranteed in the $write$s performed by the same writer. Two $write$ orders on different writers is not defined so that either one of the readers can get different order from the two $write$s. Multi-writer FIFO queue is one of the cases. In the weakest one, eventual consistency, there is no order guarantee, but the registers apply same set of $write$ operations can have the same value. Such that, readers are only guaranteed to read the same projection value at the point that every $write$ before point is delivered. Some web cache policies or storage uses this kind of policy.

Later in \ref{ch3.4}, the framework we propose shows that order of $write$s are guaranteed across passes of synchronization but not in the same pass. In the potential purpose of making writes total ordered, let each pass converged to only one $write$ left.

\subsubsection{Writing Freedom}

Writing freedom refers to the number of writer that can propose values with $write$ operations, ranging from 1 to $n$. $n$ is the total number of writers. With writing freedom, synchronization mechanisms can benefit from avoiding a specific writer to be the bottleneck of writing. Combining with leaderless approach, it enables load balancing to increase throughput, also avoids high network latency communication targets. But writing freedom can cost more rounds of communication due to coordinating concurrent operations.

\subsubsection{Latency}
To achieve certain consistency, each synchronization mechanism employs some steps. Latency is the measurement of efforts each mechanism takes on reaching the desire consistency. We use a round trip as a unit, so the latency of a synchronization mechanism is examined as the number of round trips. Some works would call it tail latency. To converge, the coordination mentioned in the above Writing Freedom section takes round trips to spread the fact. Coordination before writing value takes round trips as well. These are where latency comes from. More details will be provided in the framework. 

\subsubsection{Loading} \label{property_loading}
Loading refers to the size of the set of communication targets to spread facts. Other works may call it quorum size. In shared memory environment loading is an implicit property that is not brought to discussion in common since it is always the number of other writers. However In distributed memory environment, the synchronization mechanisms could have size ranging from 1 to $n$. $n$ is the total number of writers. Shrinking down the number of communication targets could benefit from lowering down messages counts \cite{184040}, avoiding communicating to high network latency peers \cite{10.1145/2517349.2517350} or having some spare peers just witnessing (not involving the mechanism) to perform other tasks \cite{10.1145/62546.62549}. More importantly, the size could impact the density of facts, which could lead to different fault tolerance and latency. We will have more discussion in \textsection \ref{ch:4-tradeoffs}.

\subsubsection{Fault Tolerance}
Fault tolerance is the maximum number of writer can fail at any moment during the synchronization mechanism. If more than the number of failures occur, the synchronization mechanism would not have any progress or end up with incorrect result such as split brain problem.

\subsubsection{Summary}
Combining \textsection \ref{ch3.2.1}, we can observe that there is no perfect synchronization mechanism that has all the properties at best. Aiming for both high writing freedom and low latency toward strongest consistency is impossible; Not having enough communication targets can lead to poor fault tolerance capability. Properties are trading off in different ways on each synchronization mechanism. 

\subsection{The Framework} \label{ch3.4}

We propose a framework to generalize the synchronization mechanisms. It formalizes the process of multi-writer to single-writer convergence in two ways: horizontal way and vertical way. By analyzing how synchronization mechanisms converge in different stage with spreading facts, the framework can derive properties on mechanisms and help reasoning about mechanisms.

\subsubsection{Horizontal} \label{framework_h}

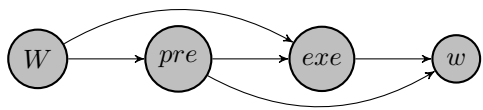
\begin{figure}[h]
\centering
\caption{A pass of synchronization.}
\label{fig:framework_h}
    \begin{tikzpicture}[node distance=2cm and 1cm,>=stealth',auto, every place/.style={draw}]
        \node [place] (S1) {$W$};
        \coordinate[node distance=1.1cm,left of=S1] (left-S1);
        \coordinate[node distance=1.1cm,right of=S1] (right-S1);
    
        \node [place] (S2) [right=of S1] {$pre$};
        \node [place] (S3) [right=of S2] {$exe$};
        \node [place] (S4) [right=of S3] {$w$};
    
        \path[->] (S1) edge  node {} (S2);
        \path[->] (S1) edge [bend left] node {} (S3);
        \path[->] (S2) edge node {} (S3);
        \path[->] (S2) [bend right] edge node {} (S4);
        \path[->] (S3) edge  node {} (S4);
    \end{tikzpicture}
\end{figure}

As mentioned in \ref{ch3.3}, a valid pass of synchronization is having writers converged and having enough writers learned the converged result to ensure safety. An executing synchronization mechanism works as consecutive passes of synchronization. Figure \ref{fig:framework_h} represents a pass of synchronization, starting from a set of $write$s issued by multiple writers to a set of $write$s that are synchronized in this pass.

$W$ represents a set of concurrent $write$ operations. Each individual $write$ operation contains a value and a metadata field, denoted as $(v, m)$. $v$ is the data intended to be written to the register and $m$ is the additional information for coordination such as (logical)timestamp, acknowledgement and dependencies. $v$ could be a null value $\phi$ or just $don't$ $care$ if $write$ is only used for spreading the fact with metadata field. $pre$ represents the stage that the synchronization mechanism is preparing for writing values. Only the metadata $m$ is used in $write$ operations in this stage. $exe$ represents the stage that the synchronization mechanism is actually executing to perform $write$ for value $v$. Metadata $m$ could be included if needed. $w$ represents a set of synchronized $write$s in this pass. The size could be one or more depending on the strength of consistency. To clarify, we use the term "stage" rather than "state" because it takes the meaning of executing different parts of the mechanism. "state" machine refers to states changing on an object, which is not the case here.

Start from the set $W$, a mechanism could converge before $pre$, in $pre$, in $exe$, or across both. Figure \ref{fig:framework_h} shows the possible ways that a pass of synchronization can walk from $W$ to $w$. There are three ways: (1) $W \to pre \to w$ (2) $W \to pre \to exe \to w$ (3) $W \to exe \to w$. During converging, writers that do not live through conflict resolution may choose to leave $write$ in $W$ for the next pass or just simply abort the operation. The one(s) that live through will then spread the converged result or merge conflicts to form the result. As mentioned right before \textsection \ref{ch3.2.1}, this is the second core aspects of synchronization mechanisms: "what to do after receiving the fact". Thus, we can observe that different synchronization mechanisms spend different round trips in different stages, with different size of $w$. This shapes each mechanism to have different properties.

For analyzing consistency, the size of converged set of $write$ indicates the consistency level. If the size is 1, the only converged $write$ is totally ordered with the previous converged $write$ and the next converged $write$. This enforces the strongest consistency level, linearizability. If the size is more than 1, the $write$s in the set is unordered, but a $write$ in current converged set is ordered with a $write$ in the previous/next converged set. For the weakest level of consistency, eventual consistency, there is no constrain on which $write$ should be in two different sets of converged $write$s since there is no order guarantee. The size of set could be any. If the registers are eventually synchronized, it means that they received the same converged set of $write$s. Projection results are then the same by applying the set. For the consistency level in-between linearizability and eventual consistency, partial order may be enforced by setting in which condition which $write$s should be in different set. For instance, if a synchronization mechanism is aiming for sequential consistency, the $write$ from the same writer should be in different sets. $write$s from different writer can either be in the same set (unordered with other $write$s) or not in the same set.

For analyzing latency, the number of round trips a synchronization mechanism spends on $pre$ and $exe$ indicates the latency to reach certain consistency level. The latency originates from the need of fact spreading and deciding what to do after receiving the fact. Again, this are the two core aspects of synchronization mechanisms. For example, the first round of Paxos (in $pre$) a writer acknowledges other writers its intention to write and retrieves the permission to write. The second round (in $exe$), assuming the permission is not preempted, the writer with permission acknowledges other registers to write the value. This shows that a pass of synchronization in Paxos spends totally two round trips, one in $pre$ and one in $exe$. Further more, the latency cost in different stage along with converging at which stage could derive different properties on each mechanism. Detailed analysis will be performed in the next section.

For analyzing writing freedom, during the $exe$ stage, the number of writer that can perform $write$ determines the writing freedom. Even though some leader-based synchronization mechanisms can have writers perform $write$ in $pre$ on attempting to change the leader metadata, the key impact on writing freedom is still in $exe$ since actually writing values happens much more often than writing metadata. It is important to note that the coordination of writing metadata is equivalent to the coordination of writing values. Latency analysis is applicable on both.

\subsubsection{Vertical} \label{framework_v}
The horizontal aspect of the framework describes the process of synchronization mechanisms. In this section, the vertical aspect describes the communication targets, size and response during each stage. 

On each round of communication issued by a writer, a writer needs to determine a set of writers to communicate. More generally, the term "quorum" is used for calling the set. Quorum could be designed to have intersection with prior and/or subsequent rounds. Writers in the intersection are arbiters. An intersection could be formed by choosing at least a majority of writers in every round to ensure overlapping or through the explicit decision of a certain writer's participation in every round. Arbiter(s) has the fact of $write$s from prior round and current round. Thus, they could be used for conflict resolution by order arbitration. The arbitration is given by certain rules in synchronization mechanisms. In response of receiving facts, the rules converge $write$s by only accepting one $write$ and acknowledging abortion to the rest, or merging a set of $write$s as a single $write$. The former guarantees the order of $write$s by linearizing them in different $w$ (figure \ref{fig:framework_h}), the later makes the $write$s that do not need order converge into the same $w$. On the other side, if arbiter(s) are not set up by having overlapping quorum or explicit decision, any order is not guaranteed for $write$s, which could eventually converge into a set of $w$.

On deciding arbiters, there are two ways, dynamic and static. Knowing the arbiter before writing values is the static way, where as only knowing the arbiter after writing values is the dynamic way. The static arbiter is decided when $write$s converged in $pre$ and the dynamic arbiter is decided during converging $write$s in $exe$. By using static arbiter, we could have quorum size at more than a half while using dynamic arbiter needs larger quorum size to prevents loosing facts between passes that caused by failures. More discussion on trading off will be discussed in \textsection \ref{ch:4-tradeoffs}.

The vertical aspect could be concluded as:
\begin{itemize} \label{list:framework_v}
    \item \textbf{Quorum:} The writers to communicate in a round. 
    \item \textbf{Arbitration:} The rules on how to response on receiving the fact.
\end{itemize}

For analyzing loading and fault tolerance, the quorum size refers the loading for a writer in a synchronization mechanism. The benefits of having lower loading in a synchronization mechanism has been mentioned in \textsection \ref{property_loading}. The size of quorum affects the density of facts. ROLL Theorem\cite{rezende2020leaderless} states that the bigger the quorum is, the synchronization mechanism could have more capability on fault tolerance. The key concept is there should be at least one writer not being faulty and having the fact of $write$s in a pass of synchronization to guarantee progress on reaching certain consistency level with correct result. Otherwise, registers could have different projection result at the end of the pass due to the split brain problem. More discussion will be in \textsection \ref{ch:4-tradeoffs}. To have more flexible approach, not only size matters. Writers could be explicitly selected instead of assuming homogeneous writers and selecting them randomly. Also quorum size could vary throughout a synchronization mechanism as long as correctness and progression of the synchronization is guaranteed. Note that some hard limits has been proposed. Such as in FLP\cite{10.1145/3149.214121}, the synchronization mechanism that proposes linearizability could not have more than a half of faulty writers. 

To summarize, table \ref{tab:my_framewrok} gives a concise coverage of the whole framework.
\begin{table}[]
    \centering
    \caption{Three steps to use the framework}
    \label{tab:my_framewrok}
    \begin{tabular}[c]{|l|l|}
        \hline
        \two{\textbf{Model}}{\textbf{Horizontally}} & Find how the mechanism walks the diagram (figure \ref{fig:framework_h}) \\
        \hline
        \two{\textbf{Model}}{\textbf{Vertically}} & State quorum and arbitration rules during each stage (list \ref{list:framework_v}) \\
        \hline
        \textbf{Analyze} & Derive properties based on the modeling result \\
        \hline
    \end{tabular}
\end{table}

\subsection{Model and Analyze Synchronization Mechanisms} \label{ch3.5}
In this section, classic synchronization mechanisms are modeled and analyzed by using the framework. Table \ref{tab:sync_comp} summarizes up results for comparison.


\subsubsection{Paxos}
\begin{table}[ht!]
    \centering
    \caption{Properties in Paxos}
    \label{tab:paxos}
    \begin{tabular}[c]{|l|l|l|l|l|}
        \hline
        \textbf{Consistency} & \textbf{W. Freedom} & \textbf{Latency} & \textbf{Loading} & \textbf{Fault Tolerance} \\
        \hline
        1  & $n$   & 2 & \two{$n$, in $pre$}{$\lceil \frac{n+1}{2}\rceil$, in $exe$}  &  $\lfloor \frac{n-1}{2} \rfloor$ \\
        \hline
    \end{tabular}
\end{table}
Horizontally, Paxos walks the path $W \to pre \to exe \to w$, with $|w| = 1$. One round trip in each stage. Vertically, Paxos takes quorum with full size in $pre$ and $\lceil \frac{n+1}{2}\rceil$ in $exe$. Although Paxos has one round trip in $pre$, it actually converges in $exe$ since every writer can still try to write value. The arbitration is performed in $exe$ by accepting the $write$ with the highest ID. Other $write$s get ignored or rejected. 

\subsubsection{Raft}
\begin{table}[ht!]
    \centering
    \caption{Properties in Raft}
    \label{tab:raft}
    \begin{tabular}[c]{|l|l|l|l|l|}
        \hline
        \textbf{Consistency} & \textbf{W. Freedom} & \textbf{Latency} & \textbf{Loading} & \textbf{Fault Tolerance} \\
        \hline
        1           & 1          & \two{1, if electing}{1, if elected} & $n$ &  $\lfloor \frac{n-1}{2} \rfloor$ \\
        \hline
    \end{tabular}
\end{table}
Horizontally, Raft walks the path $W \to pre \to w$ for electing a leader and $W \to exe \to w$ for leader writing the value, with $|w| = 1$ and one round trip in each stage. Vertically, Raft takes quorum with full size in $pre$ to converge writers to a single writer. Until the next leader change, Raft takes quorum with size $\lceil \frac{n+1}{2}\rceil$ to write value in $exe$.  

\subsubsection{Viewstamp Replication}
\begin{table}[ht!]
    \centering
    \caption{Properties in Viewstamp Replication}
    \label{tab:vr}
    \begin{tabular}[c]{|l|l|l|l|l|}
        \hline
        \textbf{Consistency} & \textbf{W. Freedom} & \textbf{Latency} & \textbf{Loading} & \textbf{Fault Tolerance} \\
        \hline
        1           & 1          & \two{1, if normal}{1.5, if changing} & $n$  &  $\lfloor \frac{n-1}{2} \rfloor$ \\
        \hline
    \end{tabular}
\end{table}
Viewstamp Replication is similar to Raft but with Fixed leader election. While Raft needs to re-elect when leader failed, VR does view change when leader failed. It walks the the same path $W \to pre \to w$ to converge the view number and $W \to exe \to w$ for leader writing the value, with $|w| = 1$. It spends one round trip to converge and half of round trip to establish authority. In later analysis we found the last half of round trip can be move to be executed asynchronously in background. Vertically, VR takes quorum with full size in $pre$ and $exe$ to converge writers and to write values.

\subsubsection{EPaxos}
\begin{table}[ht!]
    \centering
    \caption{Properties in EPaxos}
    \label{tab:epaxos}
    \begin{tabular}[c]{|l|l|l|l|l|}
        \hline
        \textbf{Consistency} & \textbf{W. Freedom} & \textbf{Latency} & \textbf{Loading} & \textbf{Fault Tolerance} \\
        \hline
        1  & $n$  & \two{1, fast}{2, slow} & \two{$\lfloor \frac{3n}{4} \rfloor$, fast}{$\lceil \frac{n+1}{2}\rceil$, slow}  &  $\lfloor \frac{n-1}{2} \rfloor$ \\
        \hline
    \end{tabular}
\end{table}
Horizontally, EPaxos walks the path $W \to exe \to w$, with $|w| = 1$. For fast path, it takes only one round trips in $exe$. For slow path, it takes two round trips in $exe$. Vertically, EPaxos takes quorum with size $\lceil \frac{3n}{4} \rceil$ in fast path and size $\lceil \frac{n+1}{2}\rceil$ in the additional round in slow path. Although it takes two round trips in slow path, EPaxos actually converges after first round trip. That is, the arbiter spreads the fact of converged writer (the first comer) to other writers when received communication. A writer will know it can commit or abort after the first found. However in EPaxos, it ensures all the $write$s will be committed so it needs an additional round to spread the dependency graph for other writers to learn the fact. If there is no conflict, second round trip is unnecessary and fast path will be taken.

\subsubsection{CRDTs}
\begin{table}[ht!]
    \centering
    \caption{Properties in CRDTs}
    \label{tab:crdt}
    \begin{tabular}[c]{|l|l|l|l|l|}
        \hline
        \textbf{Consistency} & \textbf{W. Freedom} & \textbf{Latency} & \textbf{Loading} & \textbf{Fault Tolerance} \\
        \hline
        $\mathds{Z}^+$ & $n$        &  0.5    & $[1, n]$ & $n-1$ \\
        \hline
    \end{tabular}
\end{table}
Horizontally, CRDT walks the path $W \to exe \to w$, with $|w|$ and latency could be any. Vertically, there is no quorum restriction and no arbiter needed. Besides having weaker consistency, it is hard to design as well. Otherwise it holds other properties at best.

\subsubsection{Atomic Operations}
\begin{table}[ht!]
    \centering
    \caption{Properties in Atomic Operations}
    \label{tab:atomics}
    \begin{tabular}[c]{|l|l|l|l|l|}
        \hline
        \textbf{Consistency} & \textbf{W. Freedom} & \textbf{Latency} & \textbf{Loading} & \textbf{Fault Tolerance} \\
        \hline
        1           & $n$        & 1        &   $n$      & $n-1$ \\
        \hline
    \end{tabular}
\end{table}
Horizontally, atomic operations walk the path $W \to exe \to w$, with $|w| = 1$. It spends one stage in $exe$ to check if it succeeded of failed. Vertically, since it is in shared memory environment, every access can be seen as a full size quorum broadcast. Arbitration is done by knowing if the value in the memory is changed. By this characteristic, as long as either one of the writer is not faulty, the progress is guaranteed. It is not restricted by the limits in \textsection \ref{ch:4-tradeoffs}.


\begin{table}
    \centering
    \caption{Compare properties in synchronization mechanisms.}
    \label{tab:sync_comp}
    \renewcommand{\arraystretch}{1.2}
    \begin{tabular}[t]{llllll}
        \hline
        \textbf{Mechanisms} & Consistency & W. Freedom & Latency & Loading & Fault Tolerance \\
        \hline
        \textbf{Paxos}\cite{10.1145/279227.279229}      & 1  & $n$   & 2 & \two{$n$, in $pre$}{$\lceil \frac{n+1}{2}\rceil$, in $exe$}  &  $\lfloor \frac{n-1}{2} \rfloor$      \\ 
        \hline
        \textbf{Raft}\cite{184040}       & 1           & 1          & \two{1, if electing}{1, if elected} & $n$ &  $\lfloor \frac{n-1}{2} \rfloor$       \\
        \hline
        \textbf{VR}\cite{10.1145/62546.62549}         & 1           & 1          & \two{1, if normal}{1.5, if changing} & $n$  &  $\lfloor \frac{n-1}{2} \rfloor$        \\
        \hline
        \textbf{EPaxos}\cite{10.1145/2517349.2517350}     & 1  & $n$  & \two{1, fast}{2, slow} & \two{$\lfloor \frac{3n}{4} \rfloor$, fast}{$\lceil \frac{n+1}{2}\rceil$, slow}  &  $\lfloor \frac{n-1}{2} \rfloor$       \\
        \hline
        \textbf{CRDTs}\cite{10.1007/978-3-642-24550-3_29}      & $\mathds{Z}^+$ & $n$        &  0.5    & $[1, n]$ & $n-1$          \\
        \hline
        \textbf{Atomics}    & 1           & $n$        & 1        &   $n$      & $n-1$   \\
        \hline
    \end{tabular}
\end{table}

\section{Limits and Trade-offs} \label{ch:4-tradeoffs}

In this chapter we will have a more detailed and precise analysis on how properties trade off and what are the limits. Limits are the boundary that we can not improve further. Trade-offs are the ways gaining while loosing on properties. In the previous chapter we found that there is no synchronization mechanism holds all properties at best. There is no one-size-fit-all synchronization mechanism. To summarize up the the observation, we found that:
\begin{itemize}
    \item Different level of consistency needs different level of coordination.
    \item Differences of Writing freedom enable the dynamic/static arbiter type and different coordination latency.
    \item Bigger loading trades off for better fault tolerance.
    \item Dynamic and static arbiter have different limitation on trading off loading and fault tolerance.
\end{itemize}

The analysis starts from assuming linearizability for consistency. A key problem on ensuring correctness is split brain problem. When split brain problem occurs, there will be two or more projection result at the end of a pass of synchronization. It is caused by insufficient fact spreading. Concurrent $write$s are not converged and still being concurrent. A base limit can be derived as
\begin{lemma} \label{lemma1}
    In the context of linearizability, the number of fault does not exceed $\lfloor \frac{n-1}{2} \rfloor$.
\end{lemma}
\begin{proof}
    We proof this by contradiction. Assuming failing a half of or more writers is allowed, the rule to decide static arbiter needs to accept not more than a half agreement, and the intersection of quorums to decide dynamic arbiters may not happen. Two or more static arbiters with different arbitration result can be decided from two groups of writers. Two or more dynamic arbiters with different arbitration result can be decided from two groups of writers. A pass of synchronization can then has two different projection results at the end. ($\Rightarrow\Leftarrow$)  Thus, the upper bound of fault tolerance in linearizability is $\lfloor \frac{n-1}{2} \rfloor$.
\end{proof}
This is also a result from FLP\cite{10.1145/3149.214121}. Commonly, $\lfloor \frac{n-1}{2} \rfloor$ is named as the minority and $\lfloor \frac{n+1}{2} \rfloor$ is named as the Majority. As the result, the loading can be improved by having quorum size just enough for tolerating minority faults. Larger quorum will not improve fault tolerance any further. E.g. A full size quorum will not guarantee $n-1$ fault tolerance capability.


Differences on static arbiter and dynamic arbiter originate from the writing freedom, which is the main difference between converging in different stages. Static arbiter trades the writing freedom of each writer for less latency in $exe$. Once static arbiter is decided in $pre$ with one round trip taken, until the next change of arbiter, each pass of synchronization only costs one round trip, in $exe$. Although writing freedom is possible with static arbiter by transmitting $write$ to the arbiter. Comparing to only issue $write$ from the static arbiter, it costs one extra round trip, which is not optimal if latency is concerned. For dynamic arbiter, it trades the latency for writing freedom. To ensure each $write$ is eventually executed, at least two round trips are required to complete a pass of synchronization. Notice that round trips are not required to all be in $exe$ like EPaxos. Paxos spends one round trip in $pre$ to gain permission and one round trip in $exe$ to $write$. While having writing freedom, using dynamic arbiter could suffer from the slow progression due to non-deterministic latency. Such as long dependency chain in EPaxos and bouncing back and forth between $pre$ and $exe$ in Paxos due to permission contention. To have fast synchronization mechanism with dynamic arbiter, we could abort some $write$ instead of ensure to write eventually. Fast means decide fast and abort fast. Actually this is exact what the first round trip has achieved. One round trip is enough for arbitration rule to get acknowledged and to acknowledge. Thus we have 
\begin{theorem} \label{th1}
    Under $n$ writing freedom, one round trip is optimal for deciding to write or abort. 
\end{theorem} 
\begin{proof}
    All the conflicting $write$s in a round could have arbitrary arriving order at the arbiter. By the consistency setting, only one writer can be converged to, which is the first writer. The arbitration rule determines to response negative message to all the other writers except the converged writer as positive. Thus, after one round trip each writer has the fact that it is the converged one or not.
\end{proof}
Traditionally two round trips are considered to be optimal to linearize $write$s\cite{rezende2020leaderless}. It is true for only getting facts by communication. Other local conflict resolution techniques, such as using synchronized clock, have been proposed to perform arbitration locally to reduce one round trip of communication because fact could be generated locally. As the horizontal aspect of the framework \ref{framework_h} described, $write$ metadata is equivalent to $write$ value. Latency analysis as applicable on both. Any synchronization mechanism with one round trip in $pre$ to decide static arbiter is optimal due to Theorem \ref{th1}, such as Raft. Applying it on VR, the last 0.5 round trip in $pre$ to broadcast success view change is actually is unnecessarily a part of convergence, since the view number is converged in the first round trip. So the broadcast message is the same as the heart beat message in Raft, which can be done asynchronously in background.

To decide an arbiter, either dynamic arbiter (by quorum intersection) or static arbiter (by Majority agreement), it all relies on sufficient quorum size, which is at least the Majority. The reason to check if the Majority agree on the arbiter is the same as Lemma \ref{lemma1}. It prevents two arbiters. If the synchronization mechanism design has multiple leader, it is the same that it requires the Majority agree on having those leaders. Combining the Lemma and observations, we found that it it all about quorum. As mentioned in the vertical aspect of the framework \ref{framework_v}, the key concept for loading trading for fault tolerance is that not only focus on fact spreading, but also on which writer does not get the fact. The principle is: "There must be at least one writer holding facts across two quorums to deliver them." So we can count which writers can not deliver: \textit{The writers without facts (not in the quorum)} $+$ \textit{The faulty writers}. Note that we use \textit{plus} instead of \textit{union} because a writer is either without facts or faulty, not being both at the same time. This is the key concept behind ROLL Theorem\cite{rezende2020leaderless} as well.

\begin{theorem} \label{thROLL}
    To decide a dynamic arbiter, $2(n-|Q|) + f - 2 < n$ needs to be satisfied.
\end{theorem}
\begin{proof}
    $n$ is the total number of writers, $Q$ is the quorum size and $f$ is the maximum faults tolerated. The goal is to ensure that the number of writer that can not deliver the fact to next round is less than the total number of writer. The number of the writers that in a quorum but not in the intersection is $2(n-|Q|)$, where we minimize the intersection to push to the limit. $2$ comes from having two quorums intersecting. If there are more than two quorums, either two of them should hold this theorem. The number of the faulty writers is $f$. So the number of writers that can not deliver is $2(n-|Q|) + f$. In the case that $2(n-|Q|) + f \geq n$, the writers that issue the communication quorum will know the fact that there is no writer to deliver facts. Such writers can further make larger quorum as redundancy to cover the issue as recovery strategy. Thus, the two writers that issue the quorum should be excluded from the writers that can not deliver facts, forming the limits as $2(n-|Q|) + f - 2 < n$.
\end{proof}
EPaxos uses $\lfloor \frac{3}{4}n \rfloor$ quorum size for maxing out the fault tolerance to minority, but it trades for the high loading due to big quorum size even though it as the recovery strategy. Generalized Paxos\cite{Lamport2005GeneralizedCA} has the quorum size exactly one larger than the quorum size of EPaxos due to lacking of recovery strategy. This is the same result as ROLL Theorem\cite{rezende2020leaderless}. Now we can say ROLL theorem is the theorem for fixed size quorum and dynamic arbiter situation, under the consistency level of linearizability. Notice that Lemma \ref{lemma1} still holds the maximum fault at the minority, so increasing the quorum size will not improve fault tolerance any further.

\begin{theorem} \label{thEROLL}
    To decide a static arbiter, $(n-|Q|) + f \leq \lceil \frac{n-1}{2} \rceil$ needs to be satisfied.
\end{theorem} 
\begin{proof}
    $n$ is the total number of writers, $Q$ is the quorum size and $f$ is the maximum faults tolerated. The goal is to have Majority writer have the fact of the static arbiter. The number of the writers that are not in the quorum is $n-|Q|$. The number of faulty writers is $f$. The number of writers that do not have the fact of the static arbiter is $(n-|Q|) + f$. According to Lemma \ref{lemma1}, the number should at most be the minority. Thus, the limit is form as $(n-|Q|) + f \leq \lceil \frac{n-1}{2} \rceil$.
\end{proof}
We can see Raft and VR use quorum with size of $n$ to elect leader and do view change so that they can have at most the minority failed during the process. For Paxos, some implementations use the Majority as the size of permission request quorum (the original specification allows this by not giving clear definition, or flexibility in other words). This indicates that it does not allow any fault happens during the process.

The reason to have different limits (Theorem \ref{thROLL} and Theorem \ref{thEROLL}) on static and dynamic arbiter is that the dynamic one is decided by quorum intersection and the static one is decided by receiving responses. On deciding the dynamic arbiter, the goal is to have sufficient writers stay alive after failure. While deciding the static arbiter, the goal is to have sufficient writers to fail. Yet, the key concepts behind them are the same. Notice that Paxos is categorized as using static arbiter, even though the chaining use of it creates the illusion of using dynamic arbiter. Each instance of Paxos still uses static arbiter.

Above analysis is placed for linearizability, where every $write$ should be ordered. For eventual consistency, where no ordering is guaranteed, it trades the weakest consistency for having every other properties at best. To analyze the trade-offs and limits for the synchronization mechanism with consistency in-between, simply distinguish which $write$s are commutative and categorize them as the composition of multiple series of non-commutative $write$s. Each series is treated as $write$s to be linearized. Any level of consistency can be decomposed into multiple series and to be analyzed in the known way. For example, sequential consistency can be decomposed as a series on each writer.





%



\section{Discussion} \label{ch:5-discussion}

By applying the framework, we can analyze properties on synchronization mechanisms, but notice that properties do not directly indicate performance. Properties only shape the characteristic of synchronization mechanisms. Performance should have time involved in the evaluation, either calculating time cost to gain certain property or stand alone measurement like throughput. 

Time is an interesting element in the distributed memory environment. In shared memory environment, the order of the operation can easily be determined by test-and-set, but in distributed memory environment time is not reliable due to the error of synchronized clock. Usually, logical clock\cite{10.1145/359545.359563} is used instead. The clock synchronized with Network Time Protocol typically has the error around 250ms. The synchronization mechanism must be durable to the time error to use synchronized clock. Google Spanner\cite{10.1145/2491245} shrinks down the error to about 7ms by using GPS. It makes the synchronized clock more viable. So in recent work TOQ\cite{265009}, time is proposed for conflict resolution in EPaxos. It cuts off the second round trip in slow path by applying time as local conflict resolution strategy. Local means determine order right at the writer when receiving conflicting $write$s, instead of an additional round trip of communication. This technique can also be applied to other synchronization mechanisms for reducing one round trip on conflict resolution. Since synchronized clock is not stable and reliable, this technique can only be seen as an extension, not a new mechanism. Time helps conflict resolution but it is not whole synchronization. On CRDTs side, local conflict resolution is the key to merge concurrent $write$s and having such guarantee that $write$s will eventually converged once facts are delivered.

On trading off quorum size and fault tolerance, Theorem \ref{thROLL} and Theorem \ref{thEROLL} are just an example of deriving the limits. The concept behind them can be applied to any given situation, such as dynamic quorum size or non-homogeneous quorum. In shared memory environment, data and computation are not bundled together unlike distributed memory environment, having more tolerance on computation failure. The concept can be applied to get the limit as well. On the synchronization mechanisms exploiting commutative $write$s, developer should have the awareness that originally commutative or not should be decided in application logic. Mechanisms just mitigate responsibility of developer by taking the logic into consideration.

While using the framework to analyze mechanisms, we have some extra findings. As mentioned in \ref{framework_h}, writing metadata and writing value are actually the same thing. We can apply analysis on writing each to one another. Paxos spans latency 2RTT across two stages, one in $pre$ and one in $exe$. Here we denote it as $(1, 1)$, while we can find Raft as $electing: (1,0)$ $elected: (0, 1)$ and EPaxos as $fast: (0,1)$ $slow: (0, 2)$. It can be seen as having Paxos as a general one. It can skew to $pre$ for Raft or skew to $exe$ for EPaxos. Yet, there are still more differences between them. Originally Paxos does not enforce sending response to the $write$s that are not converged to, but it is good to response to make the spread of fact faster. Fast means decide fast and abort fast. EPaxos has done this by stating conflict and having dependencies in responses, as a little sense of communication compression. EPaxos also compresses permission request and value writing in the first round trip to have the chance for fast path, where Paxos has them as two operations in two rounds. As dependency discovery in EPaxos takes the union of different dependency graph, the dependency graph can be seen as a kind of CRDT since union is commutative. Also, git is a great tool to explain all the things practically since it is a distributed version control system. All the operations such as commit, push, merge, conflict, etc. can be mapped to the behaviors in synchronization mechanisms.

\subsection{Explore New Mechanisms}
We explore new mechanisms by further improving the existing mechanism. In EPaxos revisited\cite{265009}, the evaluation shows that EPaxos suffers from high tail latency so that TOQ is proposed to solve the issue. As time is just one of the strategy to perform local conflict resolution, we found that having writers assigned with priority can also be a technique to perform local conflict resolution. Conflict operations are given orders by applying pre-assigned priority. A priority tree can be constructed as a configuration for EPaxos, having higher priority writer as parent of lower priority writer. If there are no certain priority between two writers, they are siblings. Once that the order of conflict operations, as locally determined by each author, is proven to be consistent post-application, this technique for resolving local conflicts could be deemed an equivalent solution to the TOQ approach. 

To be more specifically, the new EPaxos with priority can be modeled by the framework we proposed. Horizontally, it still walks the path as $W \to exe \to w$, with $|w| = 1$. The latency in $exe$ is reduced to 1 from 2. Vertically, quorum and arbitration rules are remained the same. Properties then can be derived as follow:

\begin{table}[ht!]
    \centering
    \caption{Improved properties in EPaxos with local conflict resolution strategy}
    \label{tab:epaxos-improved}
    \begin{tabular}[c]{|l|l|l|l|l|}
        \hline
        \textbf{Consistency} & \textbf{W. Freedom} & \textbf{Latency} & \textbf{Loading} & \textbf{Fault Tolerance} \\
        \hline
        1  & $n$  & \two{1, fast}{1, slow} & \two{$\lfloor \frac{3n}{4} \rfloor$, fast}{$\lceil \frac{n+1}{2}\rceil$, slow}  &  $\lfloor \frac{n-1}{2} \rfloor$ \\
        \hline
    \end{tabular}
\end{table}









\section{Conclusions} \label{ch:6-conclusion}

The thesis proposes a framework to generalize synchronization mechanisms from the perspective of multi-writer to single-writer convergence. By using the framework, the thesis can analyze five key properties: consistency, writing freedom, latency, loading and fault tolerance from synchronization mechanisms. It is found that there is no perfect synchronization mechanism. Properties on the mechanisms are trading off on limits. Different way of convergence in different stage with certain quorum size make each mechanism has its own characteristic. By leveraging the key concept of preventing split brain and counting which writer can not deliver facts, the thesis found the limits of properties and how they trade-off. The thesis has ROLL theorem as one of the cases in the result. The key concept can be applied on more cases for finding limits. Existed mechanisms can have further improvements such as exploiting local conflict resolution strategy.

\subsection{Future Works}
This work is only scratching the surface of synchronization. It is a first try on analyzing synchronization mechanisms generally. There are still more details to discuss. Especially on applying to real implementation. For analyzing shared memory environment, there are still more details to be covered regarding to operating system and hardware. Also in shared memory environment, the properties analyzed in the framework may not be the ones people care about. Further more, the specification of the framework could be precisely written down using TLA+ and apply to more synchronization mechanism to improve the framework.

Finally, simplicity is always what the framework insists to deliver, as the following quote can give the importance on it: \textit{"Such is modern computing: everything simple is made too complicated because it’s easy to fiddle with; everything complicated stays complicated because it’s hard to fix." -- Rob Pike (Known for Unix, Plan 9, UTF-8, Go Lang)}. In my words I would say, \textit{"Being simple is hard because you need to know all the rest are irrelevant"}



\bibliography{lipics-v2021-sample-article}

\appendix

\end{document}